\documentclass[a4paper]{jpconf} 
\usepackage{graphicx}
\usepackage{physics}
\usepackage{tensor}
\usepackage{float}
\usepackage[position=bottom]{subfig}
\usepackage[labelfont=bf]{caption}
\usepackage{bm}
\usepackage{siunitx}
\usepackage{enumitem}
\usepackage{multicol}
\usepackage{soul}
\usepackage{pgfgantt}
\usepackage{mathtools}
\usepackage{amsmath}
\usepackage{cancel}
\usepackage{enumitem}
\usepackage{etoolbox}
\usepackage{booktabs,arydshln,ltablex}
\usepackage{adjustbox}
\usepackage{multirow}
\usepackage{tikz}
\def\checkmark{\tikz\fill[scale=0.4](0,.35) -- (.25,0) -- (1,.7) -- (.25,.15) -- cycle;}
\newcommand{\Cross}{$\mathbin{\tikz [x=1.4ex,y=1.4ex,line width=.2ex] \draw (0,0) -- (1,1) (0,1) -- (1,0);}$}%

\begin{document}
\title{Physical acceptability conditions for realistic neutron star equations of state}

\author{D L Ramos-Salamanca$^{1}$, L A Núñez$^{1,2}$ and J Ospino$^{3}$}

\address{$^1$ Escuela de Física, Universidad Industrial de Santander, Bucaramanga, Colombia}
\address{$^2$ Departamento de Física, Universidad de los Andes, Mérida, Venezuela}
\address{$^3$ Departamento de Matemática Aplicada, Universidad de Salamanca, Salamanca, España}

\ead{david2208450@correo.uis.edu.co}

\begin{abstract}
We select 37 most common and realistic dense matter equation of states to integrate the general relativistic stellar structure equations for static spherically symmetric matter configurations.  For all these models, we check the compliance of the acceptability conditions that every stellar model should satisfy. It was found that some of the non-relativistic equation of states violate the causality  and/or the dominant energy condition and that adiabatic instabilities appear in the inner crust for all equation of state considered.
\end{abstract}

\section{Introduction}


Neutron stars are among the densest astronomical objects in the universe. These stars are formed from the gravitational collapse of massive stars $M > 8 M_{\odot}$ (supernova event) and leave a compact remnant whose mass and radius usually lies between $1 - 2 \,M_{\odot}$ and $10-14\,\rm{km}$, respectively \cite{HaenselPotekhinYakovlev2007}.

The inner structure of a compact object is heavily influenced by the equation of state (EoS) that governs the star's nuclear matter. This makes important theoretical predictions like the maximum mass of a neutron star (the dividing line between neutron stars and black holes) EoS dependent \cite{OzelFreire2016}. There are numerous EoS models based on models on different many-body field theories employed to describe the nuclear matter at ultra-high density \cite{FiorellaFantina2018}. These EoS do not have exact analytical form and are found in the literature as tables. Nevertheless, the EoS of neutron stars is still unknown due to the lack of a definitive theory of nuclear interactions at ultra-high densities and the impossibility of recreating the extreme density conditions in current laboratory experiments. 

Commonly, this problem involves observations to constrain individual EoS models \cite{OzelFreire2016, HernandezVivancoEtal2019}.  The present work will complement the previous methodology using physical acceptability criteria formulated in the general relativistic framework. To do this, we numerically solve the stellar structure equations for a static neutron star, using as input 37 different EoSs for the ultra-dense matter, which cover a wide variety of microscopic EoS inspired models. Subsequently, we verified that these solutions satisfied the acceptability conditions gathered by B. Ivanov \cite{Ivanov2017} and extended further by Hernández et al. \cite{HernandezNunezVasquez2018, HernandezNunezSuarez2020}. 

This paper is organized as follows. In Section \ref{stellstruct}, we describe how the relativistic equation for stellar structure are numerically solved for a given EoS. Section \ref{conditions} lists the set of conditions that any realistic stellar model (independent of the EoS) must satisfy and in Section \ref{methods} we exhibit an example for verifying these acceptability conditions. Next, in Section \ref{results}, we discuss the results of applying the physical acceptability conditions to the stellar models calculated with each EoS. Finally, in Section \ref{conclusions}, we present the conclusions of our work.

\section{Relativistic stellar structure}\label{stellstruct}
Let us assume the line element 
\begin{equation}
\dd{s}^{ 2 } = e^{ 2 \nu ( r ) } \dd{ t} ^ { 2 } - e ^ { 2 \lambda ( r ) } \dd{ r} ^ { 2 } - r ^ { 2 } \left( \dd{ \theta} ^ { 2 } + \sin ^ { 2 }  \theta  \dd{ \phi} ^ { 2 } \right),
\end{equation}
with the energy-momentum tensor given by
\begin{equation}\label{EMT}
T_\mu^\nu = \mbox{diag}\left[\rho(r),-P(r),-P(r),-P(r)  \right] \,,
\end{equation}
with $P(r)$ the pressure and $\rho(r)$ its energy density, defined in the fluid's reference frame.

The above (\ref{EMT}) leads to relativistic stellar structure equations --i.e. Tolman-Oppenheimer-Volkoff (TOV) system--, through $T^{\mu\nu}_{\quad ; \mu} =0$ , as
\begin{equation}
    \dv{m}{r} = 4\pi \rho r^2, \qquad
\frac{\mathrm{d} P}{\mathrm{d} r} = -(\rho +P)\frac{m + 4 \pi r^{3}P}{r(r-2m)} \qquad \textrm{and} \qquad \dv{\nu}{r} = - \frac{1}{\rho + P} \dv{P}{r} \,.
\label{dnutov}
\end{equation}
where $m(r)\equiv\frac{r}{2} \left(1- {\rm e}^{-2\lambda} \right)$ is defined as the gravitational mass. 
The TOV system are three first-order ordinary differential equations with four unknowns: $m(r)$, $\rho(r)$, $P(r)$ and $\nu(r)$. To close the system, we need an EoS; then we have the same number of equations and unknown functions. Thus, given $\rho_c$, the density at the central density we can integrate the system. 

\section{Physical acceptability conditions}\label{conditions}
Solutions to the Einstein field equations have to satisfy some regularity, matching and stability conditions \cite{DelgatyLake1998}. The acceptability conditions are  \cite{Ivanov2017, HernandezNunezVasquez2018, HernandezNunezSuarez2020}: {\bf C1}, Positive and free from singularities metric potentials; {\bf C2}, Matching conditions at the surface of the star; {\bf C3}, Decrease of interior redshift $Z$ with the increase of $r$; {\bf C4}, Positive density and pressures; {\bf C5}, Density and pressure having a maximum at the center and decreasing monotonically outwards; {\bf C6}, Dominant energy condition $\rho \geq P$; {\bf C7}, Causality condition $0\leq v^2 \leq 1$; {\bf C8}, Adiabatic index criterion $\Gamma > 4/3$; {\bf C9}, Stability against cracking; {\bf C10}, Harrison-Zeldovich-Novikov stability condition $dM(\rho_c)/d\rho_c>0$; {\bf C11}, Stability against convection $\rho^{\prime\prime} \leq 0$.

For the static and spherically symmetric stellar models conditions {\bf C1}-{\bf C5} and {\bf C9} are automatically satisfied. Additionally, it has been shown  \cite{Moustakidis2017} that {\bf C8} is only valid in the Newtonian limit. Thus, only {\bf C6}, {\bf C7}, {\bf C10} and {\bf C11} are relevant to our discussion. 

\section{EoS for stable models}
\label{methods}
Given a realistic EoS, $(P_i,\rho_i)$, the TOV equations are solved numerically for an evenly spaced set of central densities between $10^{14}$ g/cm$^3$ and the maximum density available in the EoS. Finally, conditions {\bf C7}, {\bf C8}, {\bf C10} and {\bf C11} are evaluated.

\begin{figure}\centering
\subfloat[Strong energy condition. The red dashed line indicates $P=\rho c^2$.]{\label{a}\includegraphics[width=.40\linewidth]{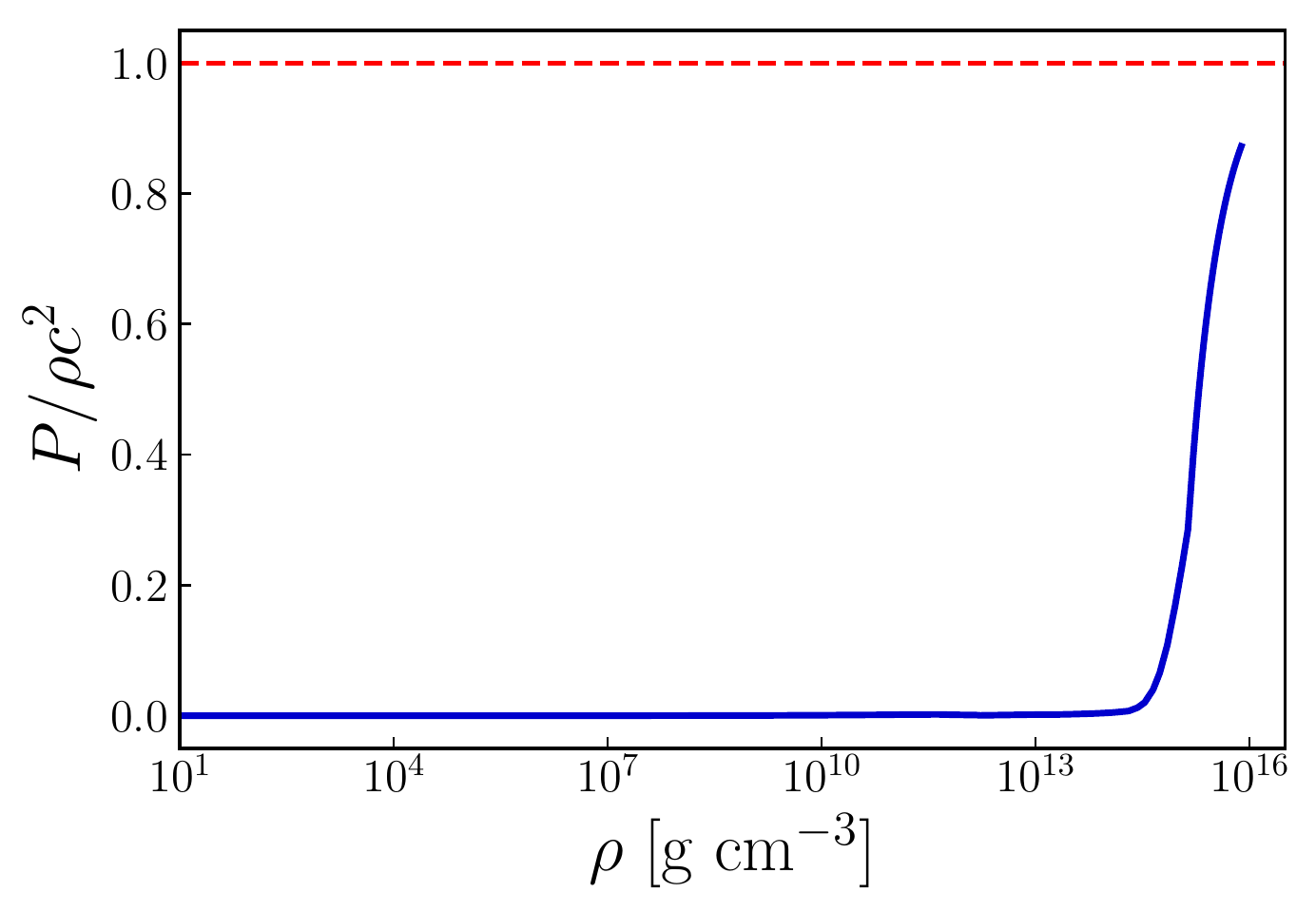}}\hfill
\subfloat[Causality condition. The red dashed line indicates $v_s=c$.]{\label{b}\includegraphics[width=.40\linewidth]{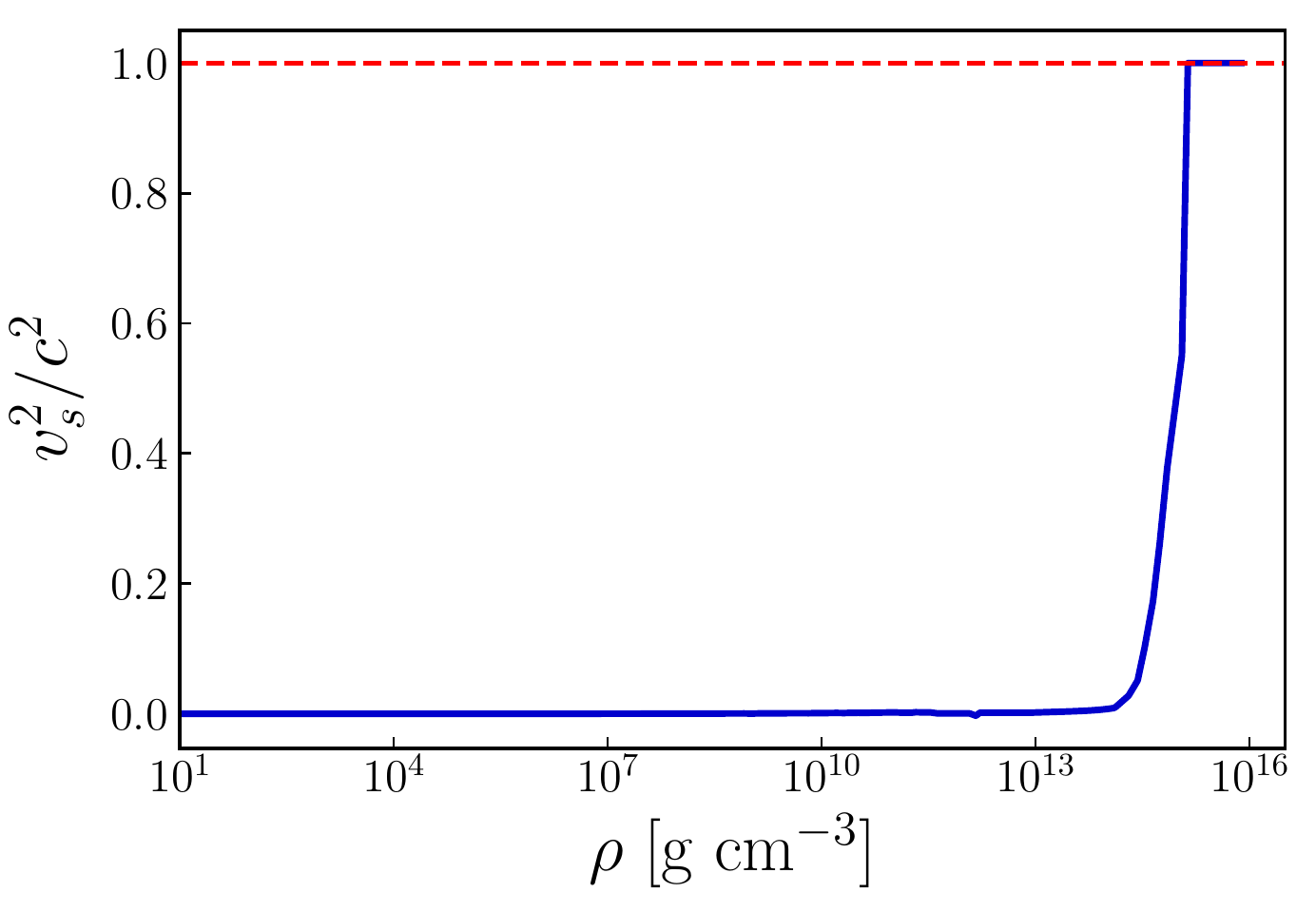}}\par 
\subfloat[$M(\rho_c)$ curve. The maximum mass for this EoS was reached for the initial value $\rho_c=2.5704 \times 10^{15}$ (red dashed line).]{\label{c}\includegraphics[width=.40\linewidth]{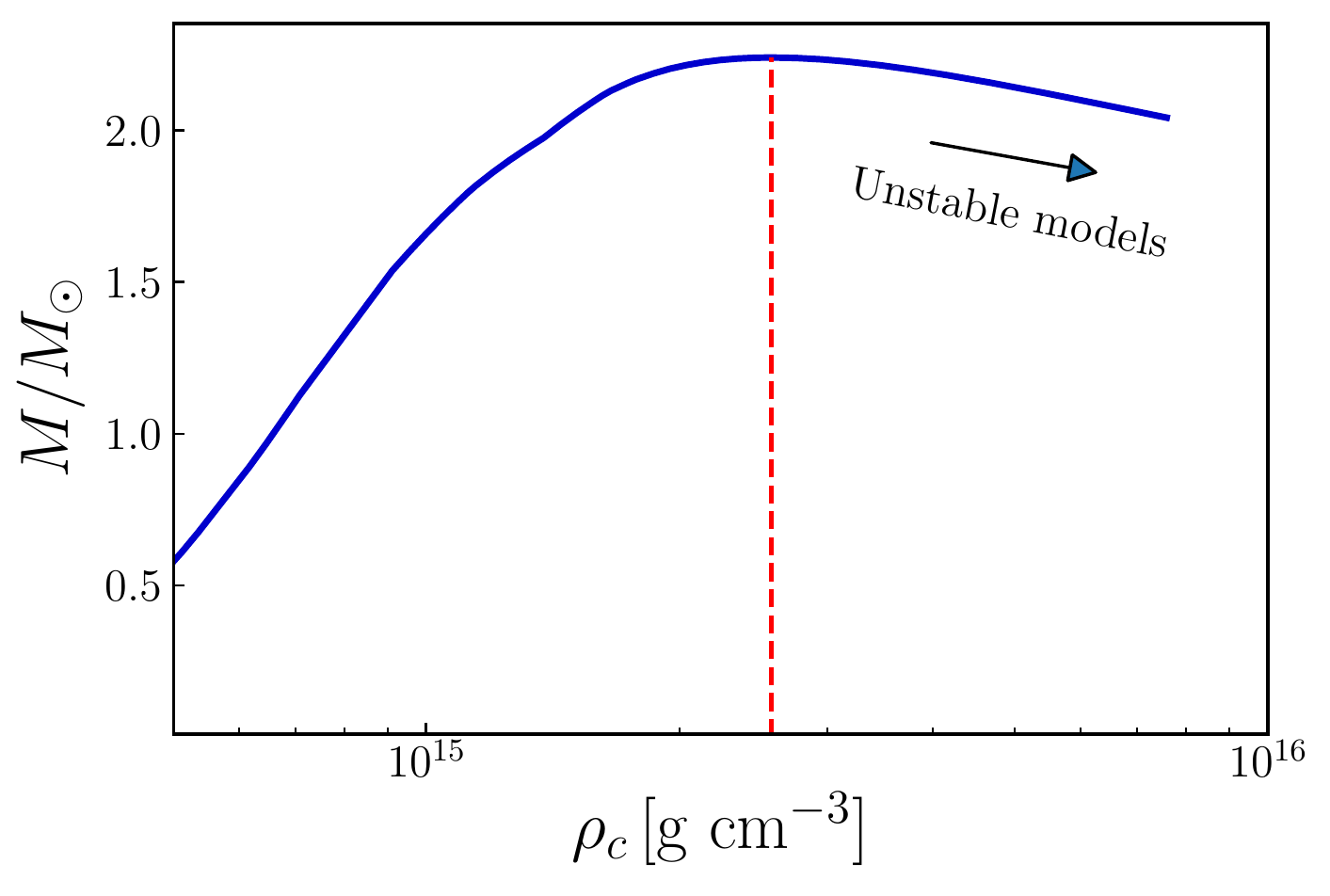}}
\caption{Verification of the \textbf{C6} (a), \textbf{C7} (b) and \textbf{C10} (c) conditions for the ENG EoS.}
\label{fig}
\end{figure}

Using the ENG EoS \cite{EngvikEtal1994} as example, conditions {\bf C6} and {\bf C7} can be checked immediately (see Figures \ref{a} and \ref{b}). Next, the TOV equations were solved with condition {\bf C10} restricting the models. Only models with $\frac { \partial M \left( \rho _ { c } \right) } { \partial \rho _ { c } } > 0$ will be considered. This derivative vanishes for the model with maximum mass and marks the start of the models that are unstable to radial pulsations (see Figure \ref{c}). 

\begin{figure}
    \centering
    \includegraphics[width=\linewidth]{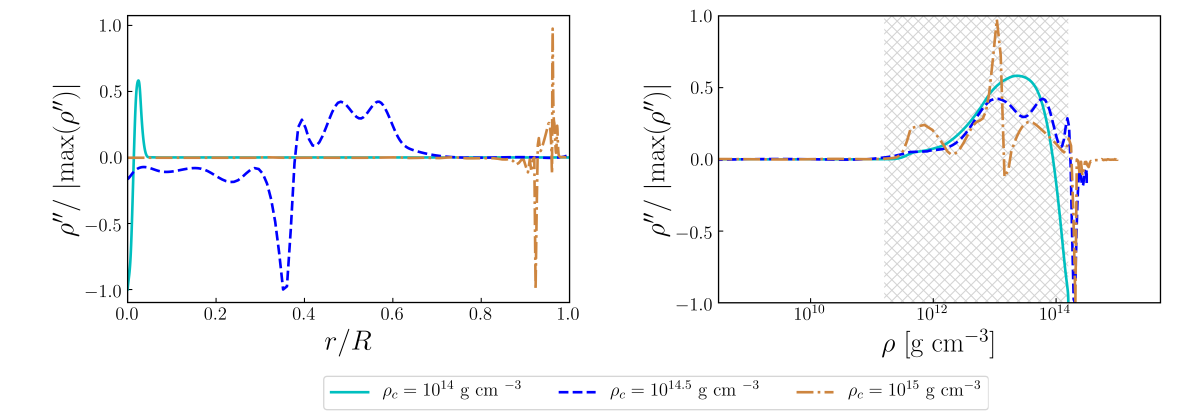}
    \caption{Convective stability condition for the ENG EoS. $\rho^{\prime\prime}(r)$ calculated for models satisfying {\bf C10}. Left. The condition does not hold over the entirety of the star. Right. It can be seen that $\rho ^{\prime \prime}>0$ in the same density range (shadowed region) for neutron stars with different $\rho_c$. This density range was found to be ($\rho_{ND},\rho_0$). }
    \label{ConvecStabilityeng}
\end{figure}

From figure \ref{ConvecStabilityeng}  it is clear that there is a density range  --from $\rho_{ND}\approx 4 \times 10^{11} $ to $\rho_0$ (shadowed)-- where {\bf C11} does not hold.  This range is consistent with the neutron star's inner crust and suggests that there could be some physical effect common to the 37 EoS considered, which generates the adiabatic instability within this region.

\section{Results and discussion}\label{results}
We analyzed a collection of 37 realistic dense matter EoS from F. Özel and P. Freire's review \cite{OzelFreire2016} and Table \ref{Consolidados} displays the results obtained. 

As reported previously \cite{HaenselPotekhinYakovlev2007}, some of the most commonly used EoSs do not comply with {\bf C6} and/or {\bf C7}. However, these violations do not occur for relativistic EoS. Thus it may suggest that the non-relativistic EoSs applied are not valid at ultra-high densities.

The results show that none of the EoSs considered produces static stellar models that are stable against adiabatic convection {\bf C11} \cite{HernandezNunezVasquez2018}. This could indicate that the non-uniformity of matter characteristic of the inner crust region can have an impact in the stability of neutron stars.

\section{Conclusions}\label{conclusions}
We consider a selection of 37 realistic dense matter EoSs --based on a wide variety of nuclear interaction models \cite{OzelFreire2016}-- to integrate the general relativistic stellar structure equations for static spherically symmetric matter configuration.  For all these models we check the compliance with the acceptability conditions {\bf C1}-{\bf C11} that every stellar model should satisfy. It was found that some of the non-relativistic EoSs infringe the causality condition and/or the dominant energy condition. 

Additionally, we obtained that adiabatic instabilities appear in the inner crust for all EoSs considered. However, this type of instability is physically challenging to interpret due to the complexity of nuclear physics in modelling this region within neutron stars.

As was previously reported in \cite{MirallesUrpinRiper1997}, the envelope of weakly magnetized stars could be unstable against convection. However, little is known about the convective stability of the inner crust in static isotropic general relativistic stars. We recognize the need to investigate further this topic, including rotation and radiation transport effects in the analysis. 

\ack
We thank the support of the Vicerrectoría de Investigación y Extensión from Universidad Industrial de Santander and the financial support from COLCIENCIAS under contract No. 8863.
 J.O. acknowledges financial support from  Ministerio de
Ciencia, Innovacion y Universidades, Spain. Grant number:
PGC2018–096038–B–I00, and Junta de Castilla y Leon, Spain.
Grant number: SA096P20.

\begin{figure}
    \centering
    \includegraphics[width=0.55\linewidth]{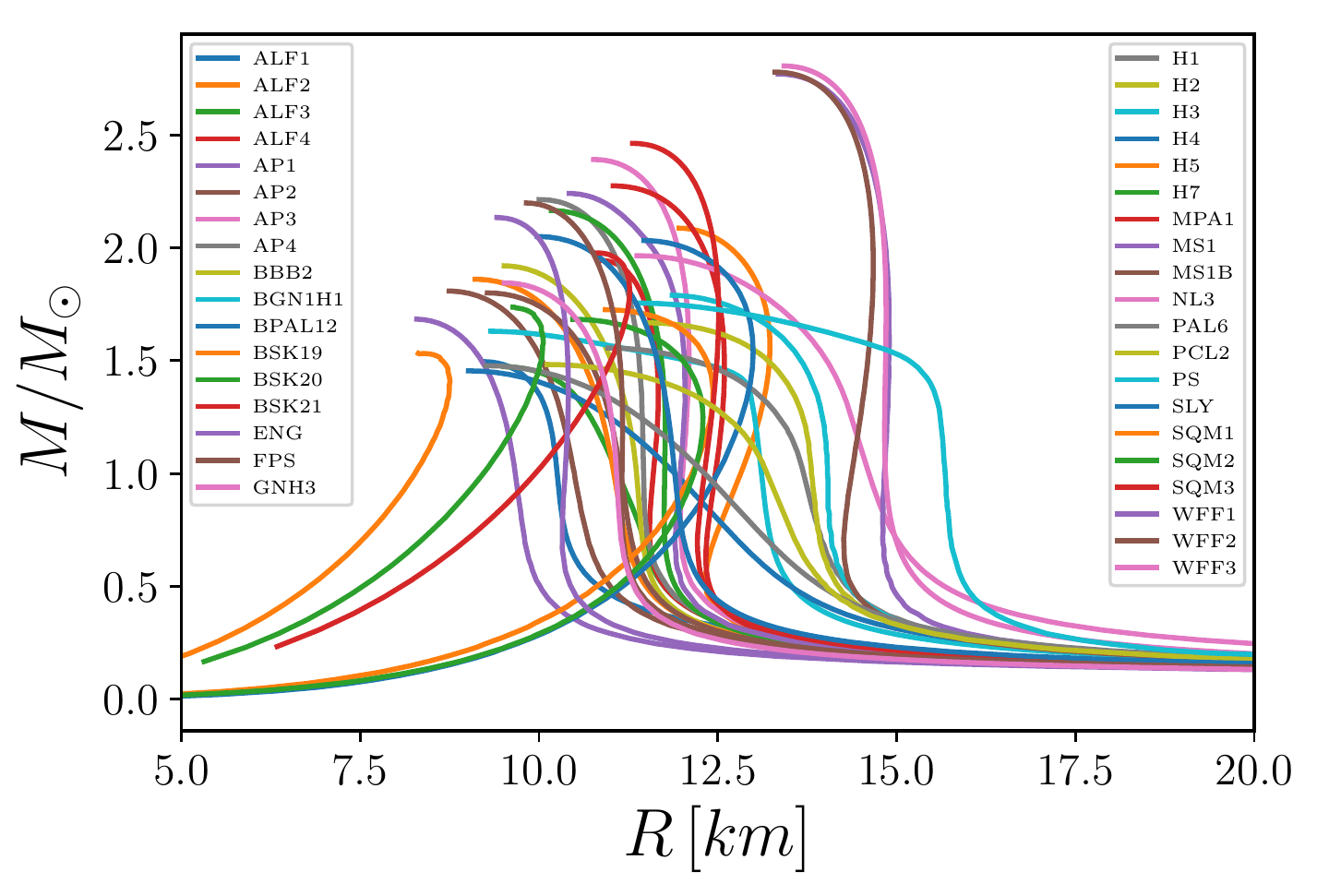}
    \caption{$M$-$R$ relations for every EoS considered.} 
    \label{MRrel}
\end{figure}

\begin{table}[H]
\caption{\small Results for the collection of 37 realistic EoSs considered. C6: dominant energy condition, C7: causality condition, C11: convection stability condition. Also shown: theoretical method used to obtain the EoS, composition (every model has leptonic contributions), maximum mass $M_{\text{max}}$ an its corresponding radius $R_{M_{\text{max}}}$ for static stars and reference to the original article.}
\label{Consolidados}
\begin{adjustbox}{max width=\textwidth}
\begin{tabular}{ccccccccccc}
\hline
\multirow{2}{*}{\textbf{EoS}} & \multirow{2}{*}{\textbf{Method}}           & \multirow{2}{*}{\textbf{Composition}} & \multirow{2}{*}{\begin{tabular}[c]{@{}c@{}}$\mathbf{M}_{\text{\textbf{max}}}$\\  $[\mathbf{M_{\odot}}]$\end{tabular}} & \multirow{2}{*}{\begin{tabular}[c]{@{}c@{}}$\mathbf{R}_{\mathbf{M}_{\text{\textbf{max}}}}$\\  $[$\textbf{km}$]$\end{tabular}} & \multirow{2}{*}{\textbf{C6}} & \multirow{2}{*}{\textbf{C7}}  & \multirow{2}{*}{\textbf{C11}} & \multirow{2}{*}{\textbf{Reference}}          \\
                     &                                   &                              &                                                                                            &                                                                                                           &                     &                                         &                      &                                      \\ \hline \addlinespace
ALF1                 & \multirow{4}{*}{Mixed}            & \multirow{4}{*}{$n,p,q$}     & 1.496                                                                                      & 9.221                                                                                              & \checkmark          & \checkmark                    & \Cross               & \multirow{4}{*}{\cite{AlfordEtal2005}}   \\
ALF2                 &                                   &                              & 2.087                                                                                      & 11.962                                                                                              & \checkmark          & \checkmark                    & \Cross               &                                      \\
ALF3                 &                                   &                              & 1.473                                                                                      & 9.514                                                                                               & \checkmark          & \checkmark                    & \Cross               &                                      \\
ALF4                 &                                   &                              & 1.943                                                                                      & 10.892                                                                                              & \checkmark          & \checkmark                    & \Cross               &                                      \\ \addlinespace
AP1                  & \multirow{4}{*}{Variational}      & \multirow{4}{*}{$n,p$}       & 1.684                                                                                      & 8.292                                                                                               & \checkmark          & \checkmark                    & \Cross               & \multirow{4}{*}{\cite{AkmalPandharipandeRavenhall1998}}    \\
AP2                  &                                   &                              & 1.809                                                                                      & 8.746                                                                                               & \checkmark          & \Cross                        & \Cross               &                                      \\
AP3                  &                                   &                              & 2.391                                                                                      & 10.765                                                                                              & \Cross              & \Cross                        & \Cross               &                                      \\
AP4                  &                                   &                              & 2.214                                                                                      & 10.004                                                                                              & \Cross              & \Cross                        & \Cross               &                                      \\ \addlinespace
BBB2                 & BD-HF                     & $n,p$                        & 1.920                                                                                      & 9.515                                                                                               & \checkmark          & \checkmark                    & \Cross               & \cite{BaldoBombaciBurgio1997}                  \\ \addlinespace
BGN1H1               & Effective potential                & $n,p,H$                      & 1.630                                                                                      & 9.325                                                                                               & \checkmark          & \checkmark                    & \Cross               & \cite{BalbergGal1997}                   \\ \addlinespace
BPAL12               & BD-HF                     & $n,p$                        & 1.455                                                                                      & 9.015                                                                                               & \checkmark          & \checkmark                    & \Cross               & \cite{ZuoBombaciLombardo1999}                       \\ \addlinespace
BSK19                & \multirow{3}{*}{Effective potential}              & \multirow{3}{*}{$n,p$}       & 1.861                                                                                      & 9.110                                                                                               & \Cross              & \Cross                        & \Cross               & \multirow{3}{*}{\cite{PotekhinEtal2013}} \\
BSK20                &                                   &                              & 2.165                                                                                      & 10.173                                                                                              & \Cross              & \Cross                        & \Cross               &                                      \\
BSK21                &                                   &                              & 2.274                                                                                      & 11.038                                                                                              & \Cross              & \Cross                        & \Cross               &                                      \\ \addlinespace
ENG                  & BD-HF                     & $n,p$                        & 2.241                                                                                      & 10.425                                                                                              & \checkmark          & \checkmark                    & \Cross               & \cite{EngvikEtal1994}                    \\ \addlinespace
FPS                  & Variational                       & $n,p$                        & 1.800                                                                                      & 9.279                                                                                               & \checkmark          & \checkmark                    & \Cross               & \cite{FriedmanPandharipande1981}                  \\ \addlinespace
GNH3                 & Field theoretical                  & $n,p,H,\Delta$               & 1.965                                                                                      & 11.372                                                                                              & \checkmark          & \checkmark                    & \Cross               & \cite{Glendenning1985}               \\ \addlinespace
H1                   & \multirow{6}{*}{Field theoretical} & \multirow{6}{*}{$n,p,H$}     & 1.556                                                                                      & 10.968                                                                                              & \checkmark          & \checkmark                    & \Cross               & \multirow{6}{*}{\cite{LackeyNayyarOwen2006}}   \\
H2                   &                                   &                              & 1.668                                                                                      & 11.516                                                                                              & \checkmark          & \checkmark                    & \Cross               &                                      \\
H3                   &                                   &                              & 1.790                                                                                      & 11.863                                                                                              & \checkmark          & \checkmark                    & \Cross               &                                      \\
H4                   &                                   &                              & 2.032                                                                                      & 11.467                                                                                              & \checkmark          & \checkmark                    & -           &                                      \\
H5                   &                                   &                              & 1.726                                                                                      & 10.930                                                                                              & \checkmark          & \checkmark                    &  -           &                                      \\
H7                   &                                   &                              & 1.683                                                                                      & 10.474                                                                                              & \checkmark          & \checkmark                    & -           &                                      \\ \addlinespace
MPA1                 & BD-HF                     & $n,p$                        & 2.462                                                                                      & 11.301                                                                                              & \checkmark          & \checkmark                    & \Cross               & \cite{MutherPrakashAinsworth1987}                    \\ \addlinespace
MS1                  & \multirow{2}{*}{Field theoretical} & \multirow{2}{*}{$n,p$}       & 2.770                                                                                      & 13.346                                                                                              & \checkmark          & \checkmark                    & \Cross               & \multirow{2}{*}{\cite{MullerSerot1996}}   \\
MS1b                 &                                   &                              & 2.778                                                                                      & 13.301                                                                                              & \checkmark          & \checkmark                    & \Cross               &                                      \\ \addlinespace
NL3                  & Field theoretical                  & $n,p,\sigma,\omega,\rho$     & 2.806                                                                                      & 13.427                                                                                              & \checkmark          & \checkmark                    & \Cross               & \cite{LalazissisKonigRing1997}                \\ \addlinespace
PAL6                 & Schematic potential                             & $n,p$                        & 1.478                                                                                      & 9.258                                                                                               & \checkmark          & \checkmark                    & \Cross               & \cite{PrakashAinsworthLattimer1988}                   \\ \addlinespace
PCL2                 & Field theoretical                  & $n,p,H,q$                    & 1.483                                                                                      & 10.116                                                                                              & \checkmark          & \checkmark                    & \Cross               & \cite{PrakashCookLattimer1995}                   \\ \addlinespace
PS                   & Field theoretical                  & $n,\pi^0$                    & 1.755                                                                                      & 11.372                                                                                              & \checkmark          & \checkmark                    & \Cross               & \cite{PandharipandeSmith1975}             \\ \addlinespace
SLy                  & Mixed                             & $n,p$                        & 2.050                                                                                      & 9.977                                                                                               & \checkmark          & \Cross                        & \Cross               & \cite{DouchinHaensel2001}                   \\ \addlinespace
SQM1                 & \multirow{3}{*}{Field theoretical} & \multirow{3}{*}{$q$}         & 1.532                                                                                      & 8.315                                                                                               & \checkmark          & \checkmark                    & -*                    & \multirow{3}{*}{\cite{PrakashCookLattimer1995}}  \\
SQM2                 &                                   &                              & 1.737                                                                                      & 9.638                                                                                               & \checkmark          & \checkmark                    & -*                    &                                      \\
SQM3                 &                                   &                              & 1.977                                                                                      & 10.814                                                                                              & \checkmark          & \checkmark                    & -*                    &                                      \\ \addlinespace
WFF1                 & \multirow{3}{*}{Variational}      & \multirow{3}{*}{$n,p$}       & 2.134                                                                                      & 9.413                                                                                               & \checkmark          & \Cross                        & \Cross               & \multirow{3}{*}{\cite{WiringaFiksFabrocini1988}}  \\
WFF2                 &                                   &                              & 2.199                                                                                      & 9.825                                                                                               & \checkmark          & \Cross                        & \Cross               &                                      \\
WFF3                 &                                   &                              & 1.845                                                                                      & 9.516                                                                                               & \checkmark          & \checkmark                    & \Cross               &                                      \\ \addlinespace

 \hline \addlinespace
\multicolumn{3}{l}{\small{*Results problematic at the star's boundary}}                                            & \multicolumn{1}{l}{}                                                                       & \multicolumn{1}{l}{}                                                                      & \multicolumn{1}{l}{} & \multicolumn{1}{l}{} & \multicolumn{1}{l}{} & \multicolumn{1}{l}{} & \multicolumn{1}{l}{} & \multicolumn{1}{l}{}  
\end{tabular}
\end{adjustbox}
\end{table}

\section*{References}
\providecommand{\newblock}{}

\end{document}